\documentclass[preprint]{aastex}

\usepackage{graphicx}
\usepackage{dcolumn}
\usepackage{amsmath,amssymb,amsfonts}
\usepackage{paralist}
\usepackage{comment}
\usepackage{graphicx,epsfig}
\usepackage{multirow}
\usepackage{wrapfig}
\usepackage{float}

\newcommand{\el}{$e^-$}
\newcommand{\sm}{$\sim$}
\newcommand{\degpd}{$^{\circ}/{\rm day}$}
\newcommand{\asps}{\arcsec/sec}

\allowdisplaybreaks

\slugcomment{To appear in Ap. J.}

\shorttitle{Finding NEA''s with Synthetic Tracking}
\shortauthors{Shao et al.}

\begin{document}

\title{Finding Very Small Near-Earth Asteroids \\using Synthetic Tracking}

\author{Michael Shao, Bijan Nemati, Chengxing Zhai, Slava G.~Turyshev, Jagmit Sandhu
}

\affil{\vskip 3pt
Jet Propulsion Laboratory, California Institute of Technology, 4800 Oak Grove Drive, Pasadena, CA 91109-0899, USA
}%

\author{Gregg Hallinan, Leon~K.~Harding
}

\affil{\vskip 3pt
California Institute of Technology, 1200 E. California Blvd., Pasadena 91125, USA
}%


\begin{abstract}

We present an approach that significantly increases the sensitivity for finding and tracking small and fast near Earth asteroids (NEA's). This approach relies on a combined use of a new generation of high-speed cameras which allow  short, high frame-rate exposures of moving objects, effectively ``freezing'' their motion, and a computationally enhanced implementation of the ``shift-and-add'' data processing technique that helps to improve the signal to noise ratio (SNR) for detection of NEA's. The SNR of a single short exposure of a dim NEA is insufficient to detect it in one frame, but by computationally searching for an appropriate velocity vector, shifting successive frames relative to each other and then co-adding the shifted frames in post-processing, we synthetically create a long-exposure image as if the telescope were tracking the object.  This approach, which we call 
``synthetic tracking,'' enhances the familiar shift-and-add technique with the ability to do a wide blind search, detect, and track dim and fast-moving NEA's in near real time. We discuss also how synthetic tracking improves the astrometry of fast moving NEA's. We apply this technique to observations of two known asteroids conducted on the Palomar 200-inch telescope and demonstrate improved SNR and 10-fold improvement of astrometric precision over the traditional long exposure approach. In the past 5 years, about 150 NEA's with absolute magnitudes H=28 ($\sim$10 m in size) or fainter have been discovered. With an upgraded version of our camera and a field of view of (28 arcmin)$^2$ on the Palomar 200-inch telescope, synthetic tracking could allow detecting up to 180 such objects per night, including very small NEAs with sizes down to 7 m.

\end{abstract}


\maketitle

\section{Introduction}

Studies of near-Earth asteroids (NEAs) have a significant value for scientific discovery, as they are believed to be remnants from the early evolution of the solar system. Accordingly, information about NEA composition and chemical properties provide important clues about conditions present during that early epoch. However, near-Earth objects also present a threat to life on Earth, as some of them may come close to and even impact the Earth. 
Notable examples include the Tunguska event over Eastern Siberia in June 1908 
\citep{Chyba1993}, 
the recent atmospheric entry of a 17 m asteroid followed by a fireball over the city of Chelyabinsk in Russia on Feb. 15, 2013 
\citep{Galimov2013}, 
and, on the same day, the flyby of a 50~m sized asteroid that passed closer to Earth than the orbits of geosynchronous satellites 
\citep{Urakawa2013}. 
In fact, a few times each year, an object of the size of a small car hits Earth's atmosphere. When these burn up on their descent through the atmosphere, they leave a beautiful trail of light known as a meteor or ``shooting star.'' Larger asteroids occasionally crash into Earth, creating craters, such as Arizona's kilometer-wide Meteor Crater near Flagstaff. Another impact site off the coast of the Yucatan Peninsula in Mexico, which is buried by ocean sediments today, is believed to be a record of the event that led to the extinction of the dinosaurs some 65 million years ago. Fortunately, these big asteroid impacts are rare.

While there has been a long history of NEA studies, the reasons above have led to recently intensified interest in the discovery and characterization of NEAs resulting in multiple currently ongoing worldwide efforts to search and catalog their population  \citep{NASA07,NRC10}. 
Two of the most productive search programs are the Panoramic Survey Telescope and Rapid Response System (Pan-STARRS)\footnote{The Panoramic Survey Telescope and Rapid Response System (Pan-STARRS):\\ {\tt http://pan-starrs.ifa.hawaii.edu/} on Maui, Hawaii \citep{Kaiser2002}, and the Catalina Sky Survey (CSS)\footnote{The Catalina Sky Survey (CSS): \\{\tt http://www.lpl.arizona.edu/css/}} in Tucson, Arizona \citep{Larson2006}.  }
To date over 10,000 NEAs have been discovered, with nearly 1,000 of these objects being over 1 km in size. In fact, the current techniques for discovering NEAs either from space or the ground have been highly successful in detecting bodies larger than 1 km.

While great strides are being made, the completeness of the known asteroid population drops rapidly for significantly smaller NEAs. In particular, the existing population of near Earth asteroids smaller than 50 m is largely under-explored \citep{Stokes03,Greestreet2012}. To date, only a small number of NEAs with sizes of 50 m have been discovered, but the vast majority, as much as 98\%, of the estimated quarter million 50 meter-class NEAs have not yet been found \citep{Har11,neo13}. Moreover, of those found, most are subsequently lost because their orbits cannot be determined with sufficient accuracy \citep{Har11}.

Recognizing the importance of finding NEAs, NASA recently announced the Asteroid Grand Challenge\footnote{The NASA Asteroid Grand Challenge: \\ {\tt http://www.nasa.gov/mission\_pages/asteroids/initiative/index.html}}, which aims to find and plan for all asteroid threats. The objectives include both planetary protection and the identification of possible human spaceflight targets for a proposed Asteroid Redirect Mission (ARM). The latter includes small NEAs with sizes 7--10 m in low delta-velocity orbits with respect to the Earth. 
 
Observing NEAs is quite different from observing the objects outside the solar system. Part of the problem is the fact that all NEAs move, even during relatively short, 30 sec exposures. In addition, small NEAs are likely to be dim (absolute magnitudes in the range H\sm28-30) and thus are observable only when they are already close to the Earth (closer than 0.1~AU). However, in that situation, the NEA is also moving very fast across the sky, with proper motion of several degrees per day (\degpd). NASA's Jet Propulsion Laboratory (JPL) maintains a website\footnote{The Near Earth Objects Program's website at JPL: {\tt http://neo.jpl.nasa.gov/}} that lists the expected upcoming \sm100 encounters with NEAs. The average velocity of these objects is \sm10 km/s relative to the Earth, which, at the distance of 0.1 AU, results in apparent proper motion of 0.14 arcsec per second (\asps). Under 1\arcsec\ seeing conditions (typical at the Palomar 200 in telescope), an exposure longer than 7 sec of the average object would result in a streaked image. Thus, traditional surveys of NEAs, with exposures of 30 seconds or more, have significantly lower sensitivity for fast moving NEAs as compared to slowly moving ones. As a result, finding and tracking of small (5-10~m in size), dim (H\sm28--30 in absolute magnitude), and fast moving NEAs (few \degpd\ in proper motion) is a major challenge.

Nearly all of the very smallest H\sm28--30 NEAs that are discovered are subsequently lost.\footnote{At the ``Target NEO 2'' workshop held on July 9, 2013, the director of the SAO Minor Planet Center was asked the question, ``What fraction of the very smallest H$\sim$28--30 NEAs that are discovered are subsequently lost?'' The answer was ``almost all of them.''}   The very smallest NEAs are detectable for only a week; at other times they are too faint to be detected. If the astrometry data collected during that week is not sufficiently accurate to predict the position of the asteroid 4--6 years in the future when it next comes close to Earth; the object is lost. This, we believe, is the great challenge in small NEA investigations at this time. Without the ability to track the NEAs after their first discovery, creating a catalog and census of these objects becomes very difficult. Similarly, planetary protection interests in these investigations require that they are not lost after initial discovery. 

Many methods are available to search and track NEAs -- all work with variable degrees of success. The ``shift and add'' technique\footnote{This method is also referred to as ``shift and stack'' and ``stack and track'' in the literature.} is a common method that is in use by many observatories for tracking Kuiper Belt objects; this technique was introduced in \cite{Tyson92}, with a first detailed description given in an HST survey \citep{Cochran1995} when it allowed for statistical detection of very faint trans-Neptunian objects. Since then the technique has been used to great success in the intervening two decades \citep{Gladman1997, Gladman1998, Gladman2001, Luu-Jewitt98, Chiang-Brown1998, Allen01, Allen02, Fraser2008, Fraser2009, Fuentes2009}. For review of the shift and add method, see \cite{Parker-Kavelaars2010} which also provides several optimized rate-sampling methods. With advent of CCD, methods of spatial-temporal \citep{Pohlig95} and maximum likelihood \citep{Sanders-Reed1998} detection, other techniques became available, leading to the matched filter trail detection technique \citep{Gural2005}, a multi-hypothesis velocity filter \citep{shucker-stuart13}.
Although, shift-and-add is a standard procedure for post-detection tracking of some NEOs (and occasional spacecraft), it has never been successfully applied to conduct a search for small, dim, and fast-moving NEA's. Partially this is because of a significant computing power requirements needed to implement this technique. This is why none of the major NEA search surveys (i.e., the CSS or Pan-STARRS) use this method for their NEO discoveries. While the NEA group at Magdalena Ridge Observatory\footnote{The Magdalena Ridge Observatory: http://www.mro.nmt.edu/} in New Mexico does use a ``track-and-stack'' technique, it is only for confirmation and follow-up purposes (same as at the Chabot Asteroid Search Program\footnote{The Chabot Asteroid Search Program:\\ http://www.chabotspace.org/asteroid-search.htm}). There are reports that the Lincoln Near-Earth Asteroid Research (LINEAR)\footnote{The Lincoln Near-Earth Asteroid Research (LINEAR):\\ http://neo.jpl.nasa.gov/programs/linear.html} \citep{Stokes2000} team experimented with track-and-stack for NEA discoveries; however, due to the anticipated large data volume and the associated significant computational efforts required, they did not adopt it as a routine procedure. We report here on a successful implementation of shift-and-add which overcomes these and other practical challenges, enabling a near real-time detection and tracking system. Hereafter we use ``synthetic tracking'' to refer to this enhanced version of the shift-and-add technique. 

In this paper, we describe why synthetic tracking can find very small NEAs with brightness down to H=30. We then show how synthetic tracking allows much improved astrometry, so that discovered objects can subsequently be tracked with enough precision to avoid being lost in the future. In section 2 we provide a brief background on detection SNR and its degradation through ``trailing loss.'' We present the details of the method along with a discussion of its performance relative to two other methods Sections 3 and 4. In Section 5, we report on test observations we made on two asteroids using the Palomar 200-inch (5.1 m) telescope on April 3, 2013 with a high-speed, low noise camera. Section 6 discusses the important topic of astrometry of NEA's and the reasons why synthetic tracking makes a significant difference in keeping track of small NEA's. Section 7 provides a closer look of the detection aspect, and the implications on the design of the major components of next-generation small-NEA detection instruments. Section 8 is a summary with conclusions.
 
\section{Background}

To detect a dim object, it is important to observe the object long enough to accumulate the necessary number of photons. The current approach to observe NEAs is to use 30-second exposures.\footnote{Pan-STARRS, for example, uses a 30 sec exposure time.} Long exposure is necessary to allow the sky noise to dominate the read noise. 
Subsequent follow up images of the same field are then recorded over some short (order 1 hour) intervals and changes in location are identified to detect moving objects.

Since small NEAs become detectable only when they are close to Earth, they typically move very fast (many \degpd\ or a few \asps) in the field. During the observable period, a long (e.g. 30 sec) exposure of such an object leaves a streak on the charged couple device (CCD) that covers many pixels instead of a single spot. 

One can estimate the loss of sensitivity from a streaked image. Many NEA searches start with separate long-exposure images of the same part of the sky taken minutes apart. The difference of any two images will remove the stationary objects, the stars. The NEA shows up as a positive blip in the first exposure and a negative blip in the second one that is subtracted. A common approach is to set the discovery threshold at 5$\sigma$, that is, a detection signal-to-noise ratio (SNR) of 5. Detection SNR is used to assess the false alarm rate and is given by the ratio of the signal over the noise from only the background. If the signal from the object is above the threshold, one could claim a detection. To avoid too many false alarms, more images are often taken. The velocity of the object must be consistent between the images for the detection to be valid. A set of three or more images is called a `tracklet' \citep{Kubica-etal_07}. If the focal plane has pixels that are smaller than the point spread function (PSF), instead of comparing each pixel against a threshold, one would perform a low pass spatial filter analysis of the image before subjecting it to a threshold. The low pass filter integrates the signal under the PSF. For reference, we will call this the `basic technique.'
In the basic technique a streaked image spreads the light across a number of pixels, reducing the signal and SNR. This loss of SNR is known as `trailing loss.' 
If the pixels are larger than the spot size, the streak penalty is lessened but, in background-limited operation, the noise level goes up. 

We can derive a simple approximate expression for trailing loss. If the transverse velocity of the object is zero, there is no streak. Over some integration time $t$ there are $S$ signal photons on top of $B$ background photons. Assuming the pixel size is matched to the (seeing-limited) width of the stellar image $w$, the background is given by $B=w^2 t \sigma_z$, where $\sigma_z$ is the `surface brightness' of the sky (zodi)  background. 
If we neglect read noise (i.e. background limited operation) then the noise is $\sqrt{B}$ so that, without a streak, the detection SNR is given by:
\begin{equation}\label{detSNRnostreak}
{\it SNR}_0 =  \frac{S}{\sqrt{B}}
\end{equation}
In general, however, the NEO will have a transverse velocity $v$ (typically many km/s) which will cause it to produce a streak of length $s=vt/r$ where $r$ is the distance to the NEO (typically a few lunar distances when first observable). This streak spans a number of pixels $(w+s)/w$ on the focal plane. 
In the basic technique, the signal is taken from the brightest pixel, so that on average the fraction of the photons in the `signal' pixel is given by:
\begin{equation}\label{Sprime}
S^\prime = \frac{w}{w+s}\,S\,.
\end{equation}
The background remains the same as the non-streaked case. This drop in the signal this causes a proportional drop in the SNR. We define trailing loss as the ratio of the SNR with smearing relative to the zero-velocity case without smearing:
\begin{equation}\label{trailoss}
\epsilon_t \equiv \frac{{\it SNR}_v}{{\it SNR}_0} = \frac{w}{w+s}\ .
\end{equation}
One might think that larger pixels that encircle more signal might mitigate the trailing loss. However, in this case one pays ``up front" by incurring a larger background and SNR loss regardless of the length of the streak. Consider the limiting case of a pixel size just large enough to enclose the entire streak. In this case there is no loss of signal, but the background is now increased by the ratio of areas. The SNR loss relative to the zero-velocity case is then simply $w/p$. While these simple equations are approximations, they serve to highlight the main effects that distinguish the different approaches to NEA searches. 

\begin{figure}
\epsscale{0.8}
\plotone{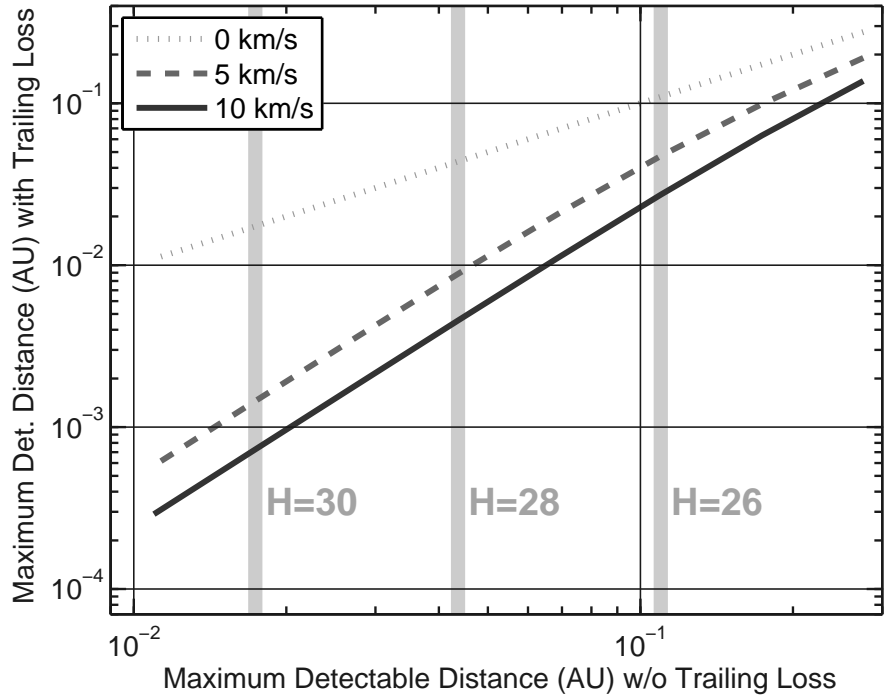}
\caption{The effect of trailing loss on the maximum detectable distance for NEAs of different H magnitudes. Plotted is the maximum detectable distance with trailing loss vs.~without trailing loss for three different relative velocities. All curves correspond to V=21.2 mag, the assumed sensitivity limit.  
\label{fig1}}
\end{figure}

Figure~\ref{fig1} shows the reduction in the distance needed to detect an asteroid when it moves at 5 km/s or 10 km/s with respect to the Earth. In this figure and throughout this paper we will assume that the limiting magnitude of a 5~m telescope is $\sim$23 mag, and the corresponding limiting magnitude for a 1~m telescope is 21.2 mag for zodi-limited detection. From this figure, which assumes a 30~s exposure, we see that an H=28 asteroid would have to be closer than 0.04 AU for its apparent magnitude to be equal to 21.2 mag. However, because of trailing losses, this object would have to be closer than 0.004 AU if its relative velocity were 10 km/s. Since volume scales as the third power of distance, a factor of 10 decrease in the distance translates into a factor 1,000 in the number of small bodies that this telescope would detect (assuming a uniform density distribution of small bodies).

\begin{figure}
\epsscale{0.8}
\plotone{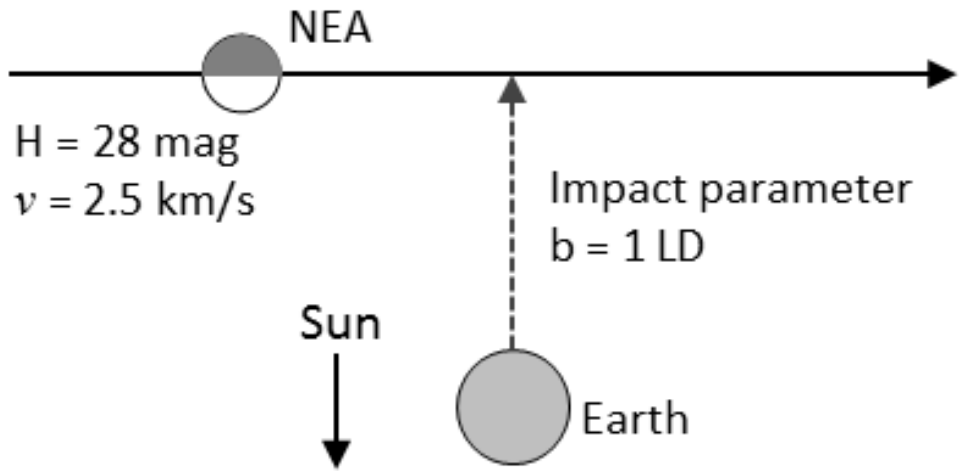}
\caption{Geometry of a simulated Earth flyby of a small, dim, and fast-moving NEA. 
\label{fig2}}
\end{figure}

\begin{figure}
\epsscale{0.8}
\plotone{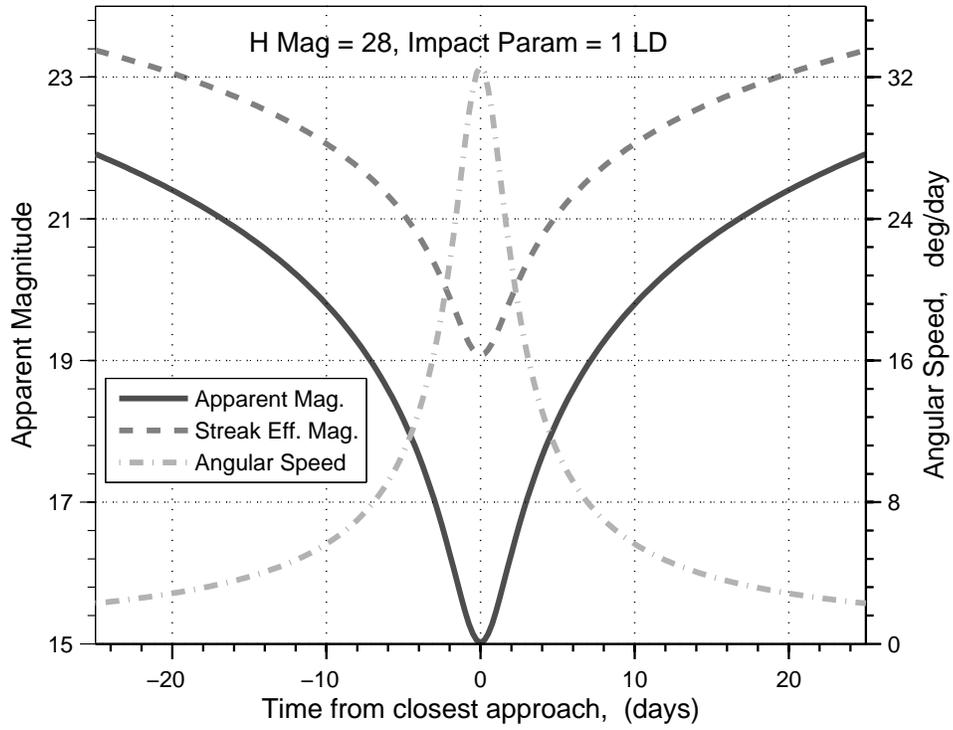}
\caption{Results of the simulated flyby shown in Fig.~\ref{fig2}. The angular velocity and apparent magnitude (with and without loss due to streaking) are shown.  
\label{fig3}}
\end{figure}

To illustrate the impact of the streak on the observable time for an NEA, we simulate a 10 m NEA of H=28 mag as it approaches the Earth with an impact parameter of 1 lunar distance (LD). This is shown in 
Fig.~\ref{fig2}. The NEA is moving parallel to the Earth's orbit around the Sun but moving 2.5 km/s faster (or slower) than the Earth. The simulation includes the phase angle effect assuming a spherical NEA and assumes Lambertian reflectance of sunlight. From this set of assumptions, in Figure~\ref{fig3} we plot a number of quantities as a function of time relative to instant of closest approach: the angular velocity on the sky, the apparent magnitude, and streak effective magnitude scaled to a 30 sec CCD exposure. This last quantity represents the loss of SNR due to the streaking of the image, and is computed using Eq.~\ref{trailoss}. It assumes the detection is based on the peak flux in a 1\arcsec\ box (the assumed seeing limit) in the streaked image.

The streaked image has two disadvantages: 1) it includes more noise from the background as it covers a larger area on the CCD; 2) the streaked shape depends on the atmospheric motion during the integration period and cannot easily be compared with a reference star in the field, leading to large errors in astrometry \citep{ver12}. Poor orbits means they will likely be lost by the time the next encounter occurs; they will be mistaken as new discoveries.

Many telescopes of different sizes are used to conduct search for NEAs. If we assume that a telescope has sufficient sensitivity to detect (at a good SNR) a stationary 21-mag star, then, at 17 days before the closest encounter, this H=28 mag object would have an apparent mag of 21. However, at 17 days, the on-sky motion would be slightly more than 3.5 \degpd, streaking the image so that the surface brightness of the streak is the same as a 22.8-mag star (Fig.~\ref{fig3}). By the time the brightness of the streaked image is brighter than 21 mag, at 5 days before closest encounter, the velocity is over 10 \degpd. Instead of being detectable for only 10 days during this encounter, synthetic tracking with the same telescope could detect this object 34 days prior to the close encounter.

\section{Synthetic Tracking Approach}

A solution to the trailing loss for fast moving objects is to use shorter exposure times such that the streak in each image is no worse than the typical 1\arcsec\ seeing limit. For an object moving at \sm20 \degpd\ (or \sm1 \asps), this suggests an exposure time of 1 sec, i.e. 1 frames per second (fps) frame rate. However, this exposure is not long enough to see faint objects. Hence, a series of exposures have to be taken, so that, when the images are subsequently shifted and added, the resultant SNR is that of a long (e.g. 30-sec) exposure. The major obstacle in this approach has been the fact that not only the traditional large format CCDs cannot be read-out at such a high rate; they would also have a prohibitively high read-noise when run at such frame rates. 

Our approach to detecting small NEAs relies on using a new type of high-speed ultra-low noise commercial-off-the-shelf (COTS) cameras originally designed for the medical imaging. These are \sm4--5 megapixel (Mpix) cameras that can take up to 100 frames/sec with only 1\el\ read noise. They represent a newer generation of CMOS technology, called Scientific CMOS (sCMOS), which enables fast frame-rate imaging. One excellent representative of this class (the Andor\texttrademark\ Zyla\footnote{For description of the Andor\texttrademark\ Zyla camera, see\\ {\tt http://www.andor.com/pdfs/literature/Andor\_sCMOS\_Brochure.pdf}}) features small pixel size (6.5~$\mu$?m), high resolution (2560$\times$2160), good dynamic range (16 bits), high frame rate (\sm100 Hz), and very importantly, low read-out noise (1.2 \el). Such a camera is well suited to the challenge of detecting faint NEAs. At a frame rate of 2 fps the read noise of the camera is lower than the zodiacal (zodi) background, thus not paying much penalty from fast reading. 

\begin{figure}
\epsscale{0.8}
\plotone{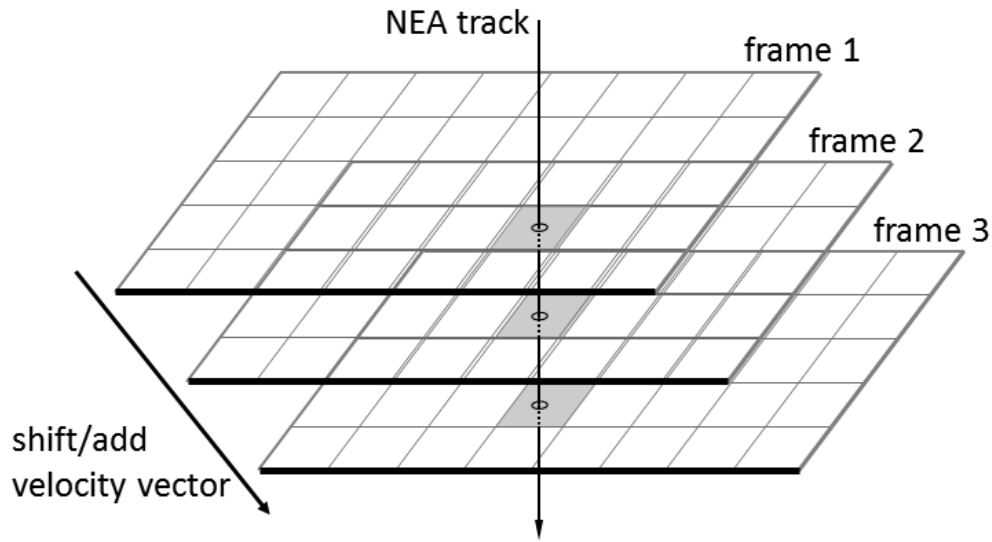}
\caption{Shift-and-add concept illustrated: because of the motion of the NEA, photons are deposited on different pixels of a CCD, but in the synthetic image (with shifted/added frames) the asteroid smear is removed. 
\label{fig4}}
\end{figure}

In the post-processing, we can shift and add these short exposure frames according to the appropriate velocity of the NEA to synthesize a long exposure, equivalent to the telescope tracking on the target. We illustrate the basic idea in Fig.~\ref{fig4}, where the vertical layers correspond to successive camera frames. Because the target moves in the field, its location in each frame is different. If we know the velocity of the NEA in advance, we can then shift the frames relative to each other according to this velocity so that the location of the NEA is kept constant with respect to the first frame. Successive images are shifted before being added. If the shift vector matches the asteroid on-sky velocity, the asteroid photons will add and the noise from the zodi background will be limited to the seeing limited PSF. As we add all the images along the vertical line (synthetic long exposure), the signal of the NEA increases linearly with the number of frames while the noise goes as the square root of the number of frames, thus SNR increases as the square root of the number of frames. With a sufficient number of frames, the NEA signal will exceed the noise from the zodi background.

A relevant example is a 1~m telescope looking for a dim, H=28 NEA. Assume the NEA has a transverse velocity of $v= 5$ km/s relative to the line of sight. Further, assume the telescope has 80\% transmission, the exposure time is $t=30$~s, the seeing limit is $w=1\arcsec$, and the zodi background is 22 mag per arcsec$^{2}$. The sky brightness is then $\sigma_z=9.6~{\rm ph}~{\rm s}^{-1}{\rm as}^{-2}$. From Fig.~\ref{fig1}, we see that this NEA is not detectable until it has approached to within $d=0.0088$~AU (3.4~LD), at which point it is at the minimum brightness of 21.2 mag for detection by this telescope and appears to move at 0.8~\asps. At 21.2 mag, the photon rate arriving at the focal plane is $r_{ph}=$20~ph/s. Using a single exposure with a camera that has quantum efficiency near  100\%, there are $r_{ph}\, t= 600$ signal photons spread across a streak that is $vt/d = 23\arcsec$ wide. The sky background under one seeing-limited pixel is $w^2\sigma_z\, t=289$ ph. The detection SNR using the basic technique is given by Eqs.~\ref{detSNRnostreak} and \ref{trailoss} and can be seen to be about 1.5. This is clearly too small to be called a detection. 

This same telescope were configured for synthetic tracking, it would use a camera that has a read noise of $n_r= 1.2\, e^-$, a quantum efficiency of $\eta=60\%$ and which is operated at a frame rate of $r_{fr} = 1$ fps (frame per second). The zodi noise in each frame adds in quadrature with the read noise. The detection SNR for this case would be:
\begin{equation}\label{STSNRdetOneMeter}
{\rm SNR}_{\rm ST}^{d} = \frac{r_{ph}\, t\, \eta}{\sqrt{r_{fr}\, t\, n_r^2+w^2\, \sigma_zt\, \eta}} = \frac{362}{14.7}=24.6 .
\end{equation}
which is approximately the same as that of the seeing limited PSF. This improvement in the SNR is similar to that reported using a matched filter technique \citep{Gural2005}.

Our technique works efficiently on detecting small and fast NEAs because the improvement of SNR by using synthetic tracking is proportional to the velocity of the asteroid in the sky. A traditional survey uses a `tracklet' (e.g. two, 30 sec exposures 15 min apart) to detect NEAs. Our analogous tracklet would be 60 images collected over 30 sec followed by an additional 60 images 15 min later. The 120 images would be analyzed as a single data set fitting the position and velocity of the object. The separation into `epochs' that are tens of minutes apart improves our efficiency for slow moving objects. For example, in a single, 30 sec epoch, objects with motion less than 1\arcsec\ per 30 sec would look stationary and not discernible from background stars. For most NEAs, requiring the motion within 30 sec to match the motion between 2 observations 15 min apart might be sufficient to claim a `discovery.' In the near future and once more data is collected, we will examine instrumental artifacts and evaluate contributing of various noise sources. This would allow us to conduct a thorough analysis of false alarms anticipated with the synthetic tracking technique. 

Before the detection of the NEA, its velocity vector is unknown.  However, we find this vector by conducting a search in velocity space. To do this we have developed an algorithm that simultaneously processes the synthetic tracking data at different velocities. The velocities searched initially have ({\it x,y}) components that are multiples of 1 pix/frame in each  direction.  This is a computationally intensive task: for example, the shift and add process for 120 images for 1,000 different velocity vectors requires over $10^{11}$ arithmetic operations. However, with current off-the-shelf graphics processing units (GPU) with up to 2,500 processors and teraFLOPS peak speeds, we were able to analyze 30 sec of data in less than 10 sec. Once the NEA is detected in this initial search, an estimate of velocity becomes possible. Using this velocity we refine the astrometry relative to a reference star in the field and determine the velocity to a much higher precision. Elsewhere we plan to describe the details of the synthetic tracking algorithm and report its performance, including its  false alarm rate. 

We now discuss setting the detection SNR threshold as well as its effect on the false positive rate and detection efficiency. The false positive rate is determined by the SNR threshold for detection, the size of CCD, and the synthetic tracking velocity space used for signal search. Assuming a Gaussian distribution for the noise in the background, using SNR>7 as the detection threshold gives a false positive probability of 1e-12 for each trial. If we search over a 2K$\times$2K image and a two-dimensional synthetic tracking velocity space with 1000 velocities, the total number of trials is then 4e9, yielding a false positive probability of 0.4\%. Setting a higher detection threshold can further reduce the false alarm rate, but would affect the detection efficiency, which is a function of the detection threshold and the signal level of the object. For example, an object with an SNR of exactly 7 would have a 50\% chance of being detected with the SNR threshold at 7, 16\% with the threshold at 8, and 84\% with the threshold at 6, where we have used the fact that the probability within a Gaussian $1\sigma$ is approximately 68\%. A trade study is needed to determine the appropriate detection threshold for any particular survey program. We note that for synthetic tracking, the detection efficiency is insensitive to the velocity of the object because there is no trailing loss. This is different from the case of using long exposure images. 

\section{Theoretical Comparison of Detection Methods}

We can quantify the SNR variation among three possible scenarios for detecting small and fast-moving NEAs, and estimate their relative performance on a `typical' H=28 NEA moving at 8 \degpd\ (0.33 \asps).

\begin{figure}
\epsscale{0.8}
\plotone{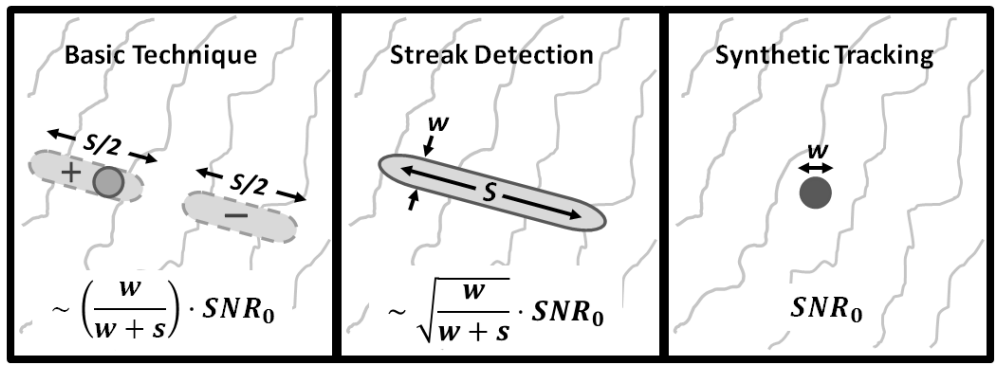}
\caption{Comparison of three methods for detecting NEA's, assuming equal exposure time.  The (relative) SNR for each of the techniques is also shown. 
\label{fig5}}
\end{figure}

These are illustrated in Fig.~\ref{fig5} and are described as follows: 
\begin{enumerate}
\item {\it Basic Technique:} Two, long-exposure (30 s) images are taken separated in time. One subtracts them and searches for a significant brightness deviations at the level of, say, 5$\sigma$. It is clear that since the NEA moves, its photons are spread out on the differential image in two streaks and peak detection sees $1/\epsilon_t$ ($\sim21$ in this case) of the NEA photons. The SNR here is 1/21 of that of a star (stationary object) of the same brightness.
\item {\it Streak Detection:}  Instead of looking for a peak in the difference of two images, one can attempt to use a matched filter to detect the streak. This method makes use of all of the photons from the NEA when the filter is a match to the actual streak. However, the larger area of the streak picks up more zodi background. The resulting SNR is $1/\sqrt{\epsilon_t}$ ($\sim0.22$) of that of a star of the same brightness, assuming the same {\em total} exposure time (60 s) as in the first case. This method could be applied to traditional surveys but, relative to the basic technique, it would be computationally intensive. The computational complexity of this technique would be comparable to synthetic tracking.
\item {\it Synthetic Tracking:} Finally, if we take short images over the same duration and then use a shift/add algorithm, we obtain the same SNR as for a star, which is a major advantage of the proposed method. To achieved this advantage, we note again that the sensor read noise must be small relative to the sky noise. 
\end{enumerate}

\section{Observation of Asteroids {\bf $2013~{\rm FQ}_{\rm 10}$} and {\bf $2009~{\rm BL}_{\rm 2}$}}

We tested the synthetic tracking approach using the CHIMERA instrument at the Palomar 200-inch telescope. CHIMERA is a high-speed, two-color photometer, newly built for the prime focus of the Palomar 200-inch.\footnote{The CHIMERA website: {\tt http://sci.esa.int/gaia/}} CHIMERA offers simultaneous observing in the Sloan g band, and either of the Sloan r or Sloan i bands. The instrument currently operates with a field of view of $2.5\arcmin\times2.5\arcmin$, with a future upgrade underway to offer an larger range of filter choices and extend the field of view to $8\arcmin\times8\arcmin$. CHIMERA uses two Andor\texttrademark\ iXon 888 EMCCD's with $1024\times1024$ pixels, allowing high-speed (10 MHz readout) imaging in two colors with very low read noise ($<$~1~\el\ with EM gain applied).

\begin{figure}
\epsscale{0.8}
\plotone{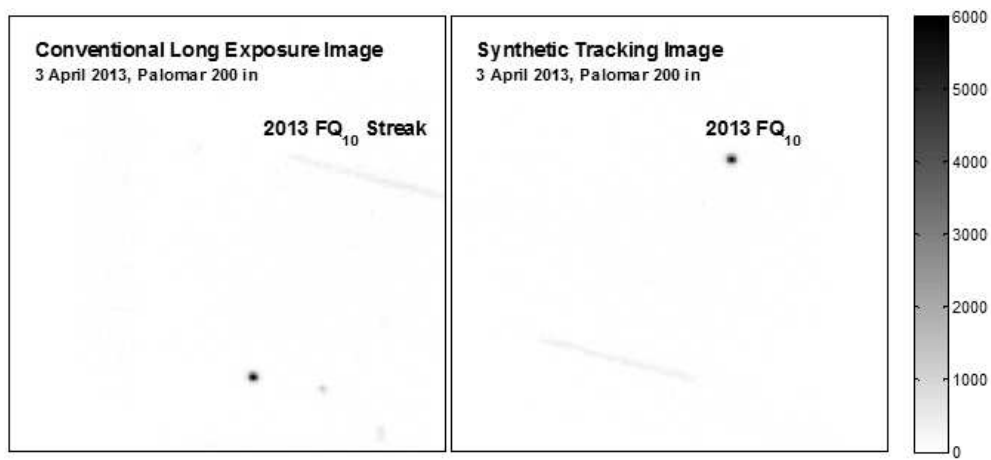}
\caption{NEA 2013FQ10 observed on April 3, 2013 at the Palomar 200 inch. On left, 700 frames are added to form a conventional long exposure. A 19-mag star, prominent on the lower part of the image, has $\sim$140,000 counts in this scale. On right, the data is processed according to the synthetic tracking approach. The asteroid now appears stationary while a faint streak is noticeable from the 19 mag star in the lower region.
\label{fig6}}
\end{figure}

On Apr 3, 2013, we observed two known NEAs, {\bf $2013~{\rm FQ}_{\rm 10}$} and {\bf $2009~{\rm BL}_{\rm 2}$}, taken from the JPL NEA list \citep{neo13}. The data was taken with an Andor\texttrademark\ iXon camera at the prime focus of the Palomar 200 inch telescope, using the Winn corrector. A Sloan $g\prime$ filter was in front of the CCD. The camera's ($512\times512$) pixels were binned ($2\times2$) on chip, resulting in a ($256\times256$)-pixel image with plate scale of 0.35 \arcsec/pixel. Data was recorded at 2 fps. The frames were time tagged with time from a GPS receiver, accurate to less than 1 msec. The results of observations of asteroid {\bf $2013~{\rm FQ}_{\rm 10}$} are shown in Fig.~\ref{fig6}(left).

{\bf $2013~{\rm FQ}_{\rm 10}$} had an apparent magnitude of \sm19.2 mag on April 3, 2013, with a velocity relative to the Earth of 9.2 km/s at a distance of \sm0.12 AU. The proper motion of the asteroid is approximately 0.1 \asps. We show 350 sec (700 frames) of data, discarding frames where the asteroid moved out of the field of view (FOV) of the camera. Fig.~\ref{fig5} shows the 700 frames co-added to create a conventional long exposure image. The streak from the asteroid is visible but faint relative to the 19-mag background star. The image uses a limited gray scale to highlight the asteroid. In this image, other background stars are visible and the 19-mag background star appears saturated. The brightness of the streak is \sm10-20 times the noise background and read noise. 

Fig.~\ref{fig6}(right) shows the result of applying a simple shift/add algorithm. The SNR of {\bf $2013~{\rm FQ}_{\rm 10}$} in the shift/add image is about 400. Here SNR is defined as the asteroid flux divided by the 1$\sigma$ noise in the background (excluding photon noise of the asteroid). Since a shift vector has been used to `freeze' the asteroid, the 19-mag star is now a streak with degraded SNR.

The noise in the background is approximately 50\% CCD read noise (\sm6\el) and 50\% photon fluctuation in the zodi background. (In the future, we plan to switch to a second generation sCMOS camera that has \sm1 \el\ read noise, although slightly lower QE.) {\bf $2013~{\rm FQ}_{\rm 10}$} has a size of \sm100 m (assuming an albedo of 12\%).  Had it been 14 m in diameter it would have been 23.4 mag and detected with a SNR of \sm8 at a distance of 0.12 AU. We saw similar results with NEA {\bf $2009~{\rm BL}_{\rm 2}$}.

\section{Astrometry}

In the last 5 years, over 150 NEAs with H\sm28--30 have been discovered, at a rate of \sm30 such objects per year \citep{neo13}. It is highly desirable to get the orbits of these objects with accuracy high enough that they will not be lost immediately after discovery. Many observatories that discover NEAs also conduct follow up observations of these objects. For bigger, \sm100 m asteroids, the follow up astrometry can be done with telescopes other than the discovering telescope weeks or even months after discovery. With 30 discoveries of 10 meter-class NEAs per year, existing `follow up' telescopes may be sufficient to get reasonable orbits. But if the discovery rate were to increase to 300 per year for small asteroids and a potentially much larger number for bigger (20--50 m class) asteroids, precise astrometry of a large number of targets may stress the existing suite of telescopes being used for follow up to get precise optical orbits.

Suggestion of the level of needed astrometric accuracy is from a simulation assuming a set of 4 astrometric measurements with \sm40 mas accuracy and taken over approximately 2 months for an NEA in a near Earth-like orbit \citep{Giorgini13}. It was found that 40 mas would be sufficient to predict the NEA position to \sm10\arcsec\ at the next apparition 3--4 years in the future. This object would not be lost. For either faster or fainter objects that conventional CCD images could not detect, synthetic tracking would be needed and the discovering facility equipped with this technology would have to also do the follow up astrometric observations. 

Ordinary CCD astrometry measures the centroids of reference stars in the frame of the CCD, as well as the position of a streaked asteroid in that 2D image. Whether the image has a streaked asteroid or a streaked star, any streaked image will result in much lower astrometric accuracy because, as we discuss below, many ``common mode'' errors such as telescope tracking errors or image motion from atmospheric turbulence will no longer be common mode.  Astrometry with synthetic tracking makes use of the 3D data set to avoid this problem by using reference stars that are bright enough so that they can be detected in a single frame.

\begin{figure}
\epsscale{0.8}
\plotone{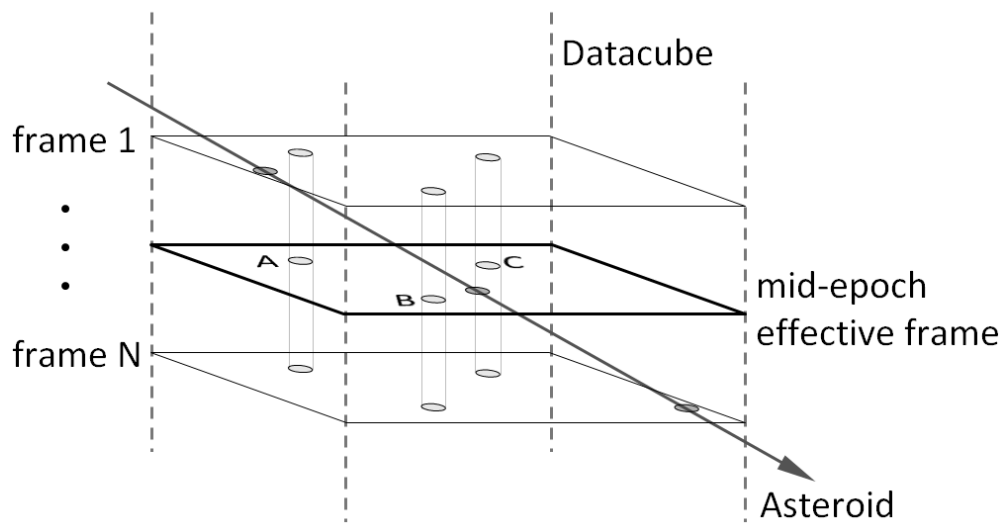}
\caption{Astrometry of asteroid using synthetic tracking. The asteroid enters the
data cube at frame 1 and exits at frame N. An effective mid-epoch frame where the asteroid and the reference stars contain all the photons from the N frames forms the basis of the astrometry.
\label{fig7}}
\end{figure}

Figure~\ref{fig7} illustrates the approach for the simple case of a telescope staring in one direction for the length of an observation (30$\sim$60~s). Synthetic tracking creates a compact image of the asteroid at an effective/virtual ``frame'' corresponding to the mid point of its travel through the data cube. The data cube consists of frames which span a certain portion of the sky, each over one slice of time. We calculate the position of the asteroid in a ``moving'' frame so that the image of the asteroid is unstreaked. We pick reference stars that are bright enough so that they can be detected in a single frame. We then average the position of the reference stars. That gives us the average position of the reference stars in the moving frame at the temporal midpoint of the data cube. 

\subsection{Astrometric Errors}

The present day optical astrometry of NEAs is much less accurate than the corresponding radar observations, sometimes by as much as 2--3 orders of magnitude. The state of the art in NEA astrometry is \sm100--200 mas 
\citep{Milani2012} \citep{Tholen 2013} whereas ground-based stellar astrometry to measure parallaxes of nearby stars can be accurate to less than 1 mas \citep{Boss09}. There are many reasons for this accuracy discrepancy. One important reason is the star catalog.  An asteroid search telescope can have a large field of view with a gigapixel focal plane composed of a mosaic of a large number of CCDs. But any astrometry of detected asteroids is limited to measuring the positions of stars within a single CCD. The spacing of the CCDs in a mosaic is neither accurate nor sufficiently stable for precision astrometry. Within the field of view of 1 CCD there are only a limited number of reference stars and this leads to lower astrometric accuracy. 

The UCAC catalog\footnote{The USNO CCD Astrograph Catalog (UCAC):\\
{\tt http://www.usno.navy.mil/USNO/astrometry/optical-IR-prod/ucac}} published by the US Naval Observatory has 50 million stars down to \sm16 mag.
Their positions are anchored to the stars in the Hipparcos and Tycho catalogs produced by the ESA Hipparcos mission, which operated in the early 1990's.  The expected accuracy of the UCAC catalog is \sm50 mas and is a major component of the \sm100--200 mas error of asteroid astrometric observations. However, in October 2013, ESA will launch the Gaia astrometric satellite\footnote{The ESA's Gaia mission's website: http://sci.esa.int/gaia/} whose very first preliminary catalog (expected \sm18 months later) will provide a reference frame better than 1 mas for objects down to \sm19--20 mag. Once we combine our approach of detecting NEAs with astrometric data from the Gaia catalog, absolute astrometry of asteroids should improve by a factor of over a 100, from the current \sm100--200 mas down to just a few mas.

\subsection{Atmospheric Errors}

In studying the data from the fast camera on Palomar, we identified an important source of astrometric error applicable to asteroid observations but not to stellar observations. It is well known that atmospheric turbulence can cause motion of the star by \sm200 mas or more. Stellar astrometry can nonetheless be done at 1-3 mas because this atmospheric error is highly correlated (`common-mode') between the stars in the field. Most atmospheric turbulence is in the lower part of the atmosphere, where the resulting image motion is common to all the stars in the field. Differential motion is a result of turbulence at the top of the atmosphere, at altitudes of \sm10 km, and is much smaller. The key to accurate differential astrometry is to make simultaneous measurements of the position of the target and reference stars. A time lag between the measurement of the target star and reference stars can result in much larger atmospheric errors. For a streaked asteroid image, the astrometric accuracy in the direction of the streak is obviously going to be worse than in the narrow direction of the streak. However, we believe a bigger effect of the streaked image is due to a non-simultaneity effect. In a 30 sec exposure, where the image is streaked by, say, 4\arcsec, the initial quarter of the streak represents the first 7.5 sec and the final quarter of the streak the last 7.5 sec. For the reference stars, on the other hand, we have the average position over the entire 30 sec exposure.
 
\begin{figure}
\epsscale{0.8}
\plotone{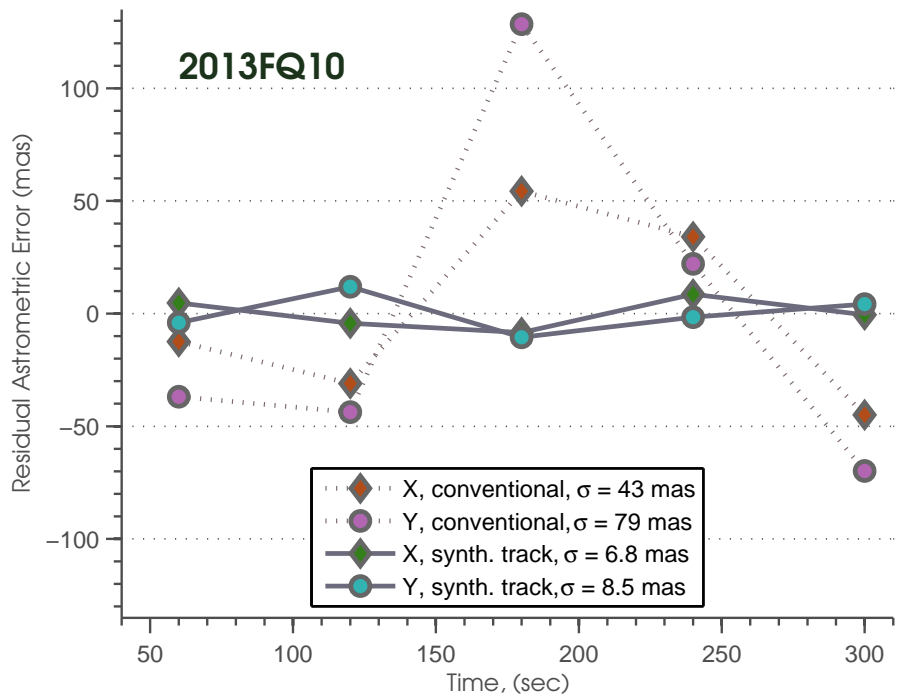}
\caption{For asteroid 2010FQ10 observed with the Palomar 200 inch telescope, the astrometric error is reduced from \sm80 mas for the streaked image (`conventional') to 9 mas for the de-streaked image using synthetic tracking.
\label{fig8}}
\end{figure}

We were able to test this hypothesis using our data on the asteroid 2013FQ10. We divided 600 frames data taken at 2 fps into 1-min segments and analyzed the data in two ways. In method 1, the frames within a 1 minute block were simply co-added to generate a conventional streaked image. In method 2, they were co-added using synthetic tracking. When the straight line motion of the asteroid is removed, the residuals are as shown in Fig.~\ref{fig8}. Astrometric residuals in the streaked image were \sm80 mas in the direction of the streak, while the astrometric residuals were \sm9 mas for the synthetically tracked images. Once the first Gaia catalog will be published, a few mas absolute astrometry of fast moving NEAs should be possible.
The astrometric error in the streaked case is relatively low, under 100 mas, because the large aperture of the telescope averages over more atmosphere than a smaller telescope. A 1-2 m telescope searching for NEAs might expect 2-3 times larger atmospheric error, perhaps up to 200 mas for long streaked images.

In a subsequent paper \citep{zhai13} we will analyze the effect of non-simultaneity in more detail.  Obviously if the streak is only 1.2\arcsec\ long, we would expect the astrometric precision to be close to a single digit milliarcsec precision of ground based stellar astronomers. However, NEAs are most often observed when they are closest to Earth. The loss of astrometric accuracy as a function of the length of the streak (in arcsec and seconds of time) will be explored and reported elsewhere. However our results on the asteroid 2013FQ10 show that the atmospheric error for a 1 minute measurement can be $< 10$ mas.  

\subsection{Photon Noise}

For the smallest NEAs, however, the limiting factor in astrometric accuracy may not be atmospheric turbulence but photon noise. In general, the precision $\alpha$ of an astrometric measurement improves with the photometric SNR according to $\alpha=w/(2\cdot {\rm SNR})$, where $w$ is the size of the spot or streak along the astrometric direction of interest. In photometric SNR, in contrast with detection SNR mentioned earlier, noise must also include the shot noise from the signal. For faint NEAs, synthetic tracking has higher photometric SNR because all the photons from the target are in a compact, seeing-limited image. As a result, for the fast moving ones synthetic tracking could detect objects normal CCD imaging cannot. It is useful to compare astrometric accuracy for an object that could be detected with normal CCD images with that from synthetic tracking. Consider an object that has a (1\arcsec$\times$4\arcsec) streak (in a nominal 30 sec exposure) with the surface brightness of the streak equal to a star with SNR of 5. If we assume the astrometry analysis uses an optimal matched filter (e.g. an elongated Gaussian PSF) that also estimates the length and orientation of the streak, then all the signal photons would be used, enhancing the SNR roughly by the ratio $\sqrt{4\arcsec/1\arcsec}$.  Along the streak, the noise-limited astrometric precision would then be $4\arcsec/(2\cdot5\cdot\sqrt{4}) \sim 0.2\arcsec$. For synthetic tracking, on the other hand, the SNR for this example would be enhanced by a factor of about 4, to \sm20, so that the noise-limited astrometric precision would be $1\arcsec/(2\cdot20)\sim 0.025\arcsec$. For dim and fast moving objects, synthetic tracking offers a potential 8-fold improvement in astrometric precision, and this is not a small factor---it means reducing the required observation time by a factor of 64. For the smallest NEAs, astrometry of the `discovery' images will be SNR limited to \sm0.1\arcsec-0.07\arcsec.

\subsection{Instrument Errors}

The astrometric error of a measurement is not just due to SNR or the atmosphere. It is the quadrature sum of both plus any instrumental errors. Because stellar parallax measurements have already demonstrated 1 mas accuracy, we expect that instrumental errors at the few milliarcsec level can be achieved. After the release of Gaia's first catalog, the atmosphere and SNR will be the major contributing terms to errors of NEAs astrometry. If one adopts an observation cadence with longer integration times to get to 30 mas accuracy, this will reduce the detection to discovery ratio from 100:1 to 1:1. We are currently conducting a relevant study, results of which will be reported elsewhere.  

\section{Design considerations for a small-NEA Search Instrument}

We now consider some general guiding principles in constructing an NEA search facility capable of searching the largely unexplored sub-100 meter NEA population, with particular attention to aspects relevant to a high-sensitivity instrument that employs synthetic tracking. 

The ability of a telescope to survey large patches of the sky is given by its \'etendue ({\it e.g.\ }\cite{Tonry11}).
 \'Etendue is defined mathematically as the product of the light collecting area A and the (solid angular) field of view of the telescope $\Omega$, or ($A\cdot\Omega$). This quantity indicates the number of photons per unit frequency per unit time a telescope will accept, and is particularly important in conducting a large astronomical survey. For a wide variety of scientific investigations, one can trade solid angle for area. A larger area reduces integration time to reach a certain limiting magnitude but a small telescope with a large field of view can achieve the same performance as a larger one with a small field of view when the goal is to measure the brightness of every object. However, for some types of observations, one cannot substitute solid angle for area. If one is looking for millisecond variability of an object and the brightest object of that type is 18 mag, there is no substitute for a large collecting area. A search for small asteroids is an application where area and solid angle are not equivalent. For detection of small asteroids, the figure of merit is the volume of space in which certain sized objects can be detected per unit of time.  

To illustrate, we compare two telescopes with the same \'etendue. The first is a 5~m telescope with a $= 0.1^\circ$ field of view using a sCMOS ($2{\rm K}\times2{\rm K}$) detector, and the second is a 1 m telescope with FOV $= 0.5^\circ$, and a CCD  with $4{\rm K}\times4{\rm K}$ pixels. We can ask: ``In a 30 sec observation, over what volume of space (in the anti-sun direction) could each of these two instruments detect an H=28 object that is moving with a transverse velocity of 10 km/s?'' Had we been surveying for stars, the two would be equivalent; the smaller telescope could take 25 exposures, 30 sec each, to make up for the smaller collecting area, while the larger telescope would spend 30 sec looking at 25 different ($0.5^\circ\times 0.5^\circ$) parts of the sky. But there are additional considerations when looking for NEAs.

For the 5 m telescope using synthetic tracking, there are no trailing losses and an H = 28~mag object can be detected at 0.1~AU. For the 1 m telescope with the larger CCD focal plane and using the basic technique, there would be a streak and trailing losses. As the NEA gets closer, the apparent magnitude will make the surface brightness of the streak increase faster than the lengthening of the streak makes it dimmer. At a distance of 0.0038 AU, the surface brightness of the streak will be equal to a 21.2 mag star (the streak will be 90$\arcsec$ long). The volume of space searched by the two telescopes would compare as follows:
\begin{equation}\label{table1}
\begin{split}
V_{5\rm m} &=  \frac{4\pi}{3}\cdot\left( 0.1~{\rm AU}\right)^3
 \cdot\frac{\left(0.1\deg\right)^2}{41253\deg^2} =1.0\times10^{-9}~{\rm AU}^3,  
 \\
V_{1\rm m} &=  \frac{4\pi}{3}\cdot\left( 0.0046~{\rm AU}\right)^3
 \cdot\frac{\left(0.5\deg\right)^2}{41253\deg^2} =2.4\times10^{-12}~{\rm AU}^3.
 \end{split}
\end{equation}
The volume of space surveyed by the 5-meter telescope with synthetic tracking is over 400 times larger, both systems having identical ($A\cdot\Omega$). While the example helps to illustrate that $A\cdot\Omega$ is not the whole story, one would not in practice equip the 1-meter  conventional telescope in this way. We will next start from a more realistic reference case with a wider field and consider a progression of more optimized facilities from the standpoint of accessible search volume.

Since the 1-meter telescope can support a much wider field, we replace the focal plane with a $10{\rm K}\times10{\rm K}$ ($9\mu{m}$) CCD and add a lens system that brings the focal ratio of the telescope to $f/2.1$, expanding the field of view to  $1.5^\circ$. This stretches the accessible search volume by a factor of  9 relative to the 1-meter example above, making it a more realistic reference point for the further improvements now described and summarized in Table~\ref{TelCompare}. 

If instead of the wide field CCD we furnished the 1-meter telescope with the fast sCMOS sensor of the 5-meter telescope and used synthetic tracking, the widest field supportable would be  $0.3^\circ$. But, as Table~\ref{TelCompare} shows, this 1-meter would now perform almost as well as the 5-meter with a 35-fold increase in searchable volume over the reference (wide-field) 1-meter telescope employing the basic technique. That the 5-meter telescope can not do better with this sensor is due to the small pixel size of the sCMOS detector; it is not practical to have the focal ratio for the 5-meter telescope be significantly faster than $f/1.5$. As a result, the small pixels grossly over-sample the 1\arcsec\ PSF.  A next-generation sCMOS camera with $4{\rm K}\times4{\rm K}$, 16~$\mu{\rm m}$ pixels would allow the 5-meter telescope to have a  $28\arcmin$ FOV with a volume advantage of over 1000 relative to the reference 1-meter.

Using a model of the distribution of NEA's \citep{Chodas13}, we have estimated the annual yields for the telescope configurations listed in Table~\ref{TelCompare} for a detection threshold of SNR=7. We have assumed 8~hours of observing time per night and  efficiency losses of 25\% for weather, 25\% for the moon, 10\% for follow-up observations. The results appear in Table~\ref{Yields}. From these results we conclude that a 5~m telescope  employing  synthetic tracking with a next-generation sCMOS sensor will have an H$>$28 NEA yield of \sm5,500 per year if it is used for 2 months of the year. As we noted in Section~1, historically the annual discovery rate of NEAs with 
H$>$28 has been \sm30 per year.


\section{Conclusions}

Large near-Earth asteroids 250 m in diameter can be detected 0.25 AU from the Earth and their motion across the sky can be slow enough that \sm30 sec CCD exposures only result in short streaks.  
However, an NEA 10 times smaller in diameter has to come 10 times closer to Earth to be detected and on average it will be moving 10 times faster across the sky. 
As the NEA moves faster across the sky, the sensitivity of conventional 30~sec CCD exposures decreases.  
The loss of sensitivity, while not very serious for a 250 m NEA, is a major factor for 10~m sized NEAs.  
This loss of sensitivity can be effectively eliminated by an implentation of the ``stack and track'' method that uses a new generation of low read-noise, high-speed sCMOS cameras and modern GPU's capable of searching a large velocity space in near-real time. 
Allowing for telescope availability of 2 months per year and observing inefficiencies, such a system implemented on a large, 5~m telescope should be able to discover \sm5500 NEAs of H=28--31 in one year, a rate which is almost 200 times higher than the discovery rate of these small objects over the last 5 years.

The other advantage of synthetic tracking is that it enables significantly higher astrometric accuracy of NEAs. 
Narrow angle astrometry with \sm5-10 mas accuracy coupled with a catalog of reference stars from the Gaia mission should mean that observation of even 10~m NEA at \sm3-4 epochs over a few days would be sufficient to measure the orbit so these objects are not subsequently lost. 
Last, milliarcsec level astrometry will enable astronomers to measure the area/mass ratio of all asteroids, and when their physical size is measured with radar or thermal IR photometry, one can calculate the masses of these NEAs.

\acknowledgments

The authors wish to thank J. Giorgini of JPL for the simulation of what astrometric accuracy is needed to ``not lose'' an asteroid in a near earth orbit, and P. Chodas of JPL for supplying the estimates for the NEA population in the range of magnitudes H$\sim$26-31. We thank V.~E. Zharov of the Lomonosov Moscow State University, A.~H. Parker of UCB,
G. McKeegan of Chabot Observatory and J. Scott Stuart of MIT for useful comments on the manuscript. We also thank the anonymous referee for a set of valuable comments that helped to improve the manuscript. The work described here was carried out at the Jet Propulsion Laboratory, California Institute of Technology, under a contract with the National Aeronautics and Space Administration. Copyright 2013 California Institute of Technology. Government sponsorship acknowledged.

\clearpage

\begin{table}
\begin{center}
\caption{Comparison of different telescope configurations and their relative search volumes, normalized to the search volume of the first row, a wide-field conventional 1~m telescope emplying the basic technique.\label{TelCompare}}
\begin{tabular}{cccccccc}
\tableline\tableline
Diam. & FOV & Technique & FPA & ${\rm N}_{\rm pix}$ & Pixel  & F\# &
Volume\\
\tableline
1~m &$1.5^\circ$ &Conventional & CCD &
	$10{\rm K}\times10{\rm K}$ & $9~{\mu\rm m}/0.54\arcsec$ &3.5&1 \\
1~m &$0.3^\circ$ &Synth. Trk.& sCMOS &
	$2{\rm K}\times2{\rm K}$ & $6.5~{\mu\rm m}/0.54\arcsec$&2.5& 35 \\
5~m &$0.1^\circ$ &Synth. Trk.& sCMOS &
	$2{\rm K}\times2{\rm K}$   & $6.5~{\mu\rm m}/0.18\arcsec$&1.5&47 \\
1~m &$0.6^\circ$ & Synth. Trk.   & \multicolumn{1}{c}{sCMOS\tablenotemark{a}}  & $4{\rm K}\times4{\rm K}$ & $16~{\mu\rm m}/0.54\arcsec$ &6.1 & 141 \\
5~m &$28\arcmin$ & Synth. Trk.   & \multicolumn{1}{c}{sCMOS\tablenotemark{a}}  & $4{\rm K}\times4{\rm K}$ & $16~{\mu\rm m}/0.42\arcsec$ &1.6 & 1023 \\
\tableline
\end{tabular}
\tablenotetext{a}{Anticipated next generation sCMOS.}
\end{center}
\end{table}

\begin{table}
\begin{center}
\caption{Expected annual yields of  NEA's with ${\rm SNR} \ge 7$ for the telescope configurations listed in Table~\ref{TelCompare}, in the  same order. The calculations assume a year-round available facility, operated 8 hours per night, with efficiency losses of: 25\% for the moon, 25\% for weather, and 10\% for follow-up observations.\label{Yields}}
\begin{tabular}{cccccc}
\tableline\tableline
Diam. & FOV & Technique & FPA & H=25--31 & H=28--31 \\
\tableline
1~m & $1.5^\circ$ & Conventional & $10{\rm K}\times10{\rm K}$ & 1,400  & 35  \\
1~m & $0.3^\circ$ & Synth. Trk.  & $2{\rm K}\times2{\rm K}  $ & 2,600  & 1,000  \\
5~m & $0.1^\circ$ & Synth. Trk.  & $2{\rm K}\times2{\rm K}  $ & 4,000  & 1,500 \\
1~m & $0.6^\circ$ & Synth. Trk.  & $4{\rm K}\times4{\rm K}  $ & 10,400 & 3,900  \\
5~m & $28\arcmin$ & Synth. Trk.  & $4{\rm K}\times4{\rm K}  $ & 86,700 & 33,000\\
\tableline
\end{tabular}
\end{center}
\end{table}


\clearpage

\end{document}